\newcommand{\qir}{\mbox {$ q_{1}^{\rho}$}}

\newcommand{\al}{\mbox{$\alpha $}}

\newcommand{\e}{\mbox{$e^{ik_{0}X}$}}
\newcommand{\qe}{\mbox{$e^{iq_{0}X}$}}

\documentstyle[12pt]{article}
\newcommand{\kim}{\mbox {$ k_{1}^{\mu}$}}

\newcommand{\ki}{\mbox {$ k_{1}$}}

\newcommand{\qi}{\mbox {$ q_{1}$}}

\newcommand{\qo}{\mbox {$ q_{0}$}}
\newcommand{\ko}{\mbox {$ k_{0}$}}

\newcommand{\be}{\begin{equation}}
\newcommand{\br}{\begin{eqnarray}}
\newcommand{\ee}{\end{equation}}
\newcommand{\er}{\end{eqnarray}}

\newcommand{\p}{\mbox {$ \partial$}}

\newcommand{\mup}{\mbox {$ \partial _{\mu}$}}
\newcommand{\nup}{\mbox {$ \partial _{\nu}$}}

\begin{document}
\title{A Comparison of the Proper Time Equation and the Renormalization
Group $\beta$-Function in String Theory}
\author{B. Sathiapalan\\ {\em
Physics Department}\\{\em Penn State University}\\{\em 120
Ridge View Drive}\\{\em Dunmore, PA 18512}}
\maketitle
\begin{abstract}
It is known that there is a proportionality factor relating the
$\beta$-function and the equations of motion viz. the Zamolodchikov
metric.  Usually this factor has to be obtained by other methods.
The proper time equation, on the other hand, is the full equation
of motion.  We explain the reasons for this and illustrate it
by calculating corrections to Maxwell's equation.
The corrections are calculated to cubic order in the field strength,
but are exact to all orders in derivatives.
We also test the gauge covariance of the
proper time method by calculating higher (covariant) derivative
corrections
to the Yang-Mills equation.
\end{abstract}
\newpage
\section{Introduction}

The $\beta$-function or renormalization group approach provides a
convenient way of obtaining the low energy equations of string theory.
\cite{l,cc,as,ft,cz}.  There have been many generalizations of this
method subsequently, to massive modes \cite{ds,ao,p,bspt,bslv}, to higher
order corrections in gauge theories \cite{ac,ft1,t,do},
to connections with
Wilson's renormalization group equations \cite{p,bm,bspt},
to connections
with string field theory \cite{bsos,ft2,t2}, to
background independent formulations \cite{w,wl,s} and in other
ways.  The proper-time approach
\cite{bspt}
is a generalization of a technique
that has been used for point particles\cite{rf,js,yn,pm,sm,ct}
and is a variant of the $\beta$-function
approach that has some advantages. One
of the advantages is that by virtue of its similarity with an S-matrix
calculation it is guaranteed to
reproduce the full equations of motion. The $\beta$-function,
on the other hand,
is not identical to the equation of motion -
it is only {\em proportional} to the equation of motion.
This was shown in the case of the tachyon \cite{bspt}.

More recently
this method has been generalized to include gauge fields \cite{bsg,bsfc},
both, Abelian and non-Abelian.  In the Abelian case we derived a string
generalization of the covariant Klein Gordon equation for scalar
particles
coupled to electromagnetism.
In the non-Abelian case we derived the Yang Mills equation.  Some
time ago it was shown that string corrections to Maxwell action
gives the Born-Infeld action in a low energy limit \cite{ft1,t}.  This
was done by calculating the
partition function in the presence of background
electromagnetic fields.  Subsequently it was shown that the same result
could be obtained in the conventional
$\beta$-function approach \cite{ac}.
However it was found that the $\beta$-function is not quite equal to the
equations of motion, rather it was proportional to it\cite{ac}.
The prefactor was fixed by requiring that the equations
come from an action.
This
proportionality factor is the Zamolodchikov metric\cite{zam} as was shown
in ref\cite{polya}.
Polyakov showed that the equations of motion are related to
the $\beta $ function as follows:
\be
\frac{\delta {\cal L}}{\delta \phi ^{i}} =
G_{ij} \beta ^{j}
\ee
Here $G_{ij}$ is the Zamolodchikov metric.
This was later worked out for
the case of the tachyon
\cite{bspt}.  As mentioned earlier, it was also shown there
that the proper time
equation gives the full equation, prefactor and all.  It is a natural
question,then, as to whether one can get the full corrections
(including the Zamolodchicov metric)
to the Maxwell
equation using the proper time approach.
          In this paper we address this question and show that
the proper time equation does give the full equation of motion.
We do this in two different ways.  The first method is a straightforward
generalization of \cite{bspt} and
uses on-shell gauge fixed vertex operators.
This calculation is very similar to an S-matrix calculation.  Thus we
obtain the leading (non-trivial) term of the Born-Infeld action.  But in
addition it has all the momentum dependence included,i.e.,all the
higher derivative corrections to the low energy equations of motion
are included.
  In this sense
it goes beyond the conventional $\beta$-function calculation.
The second method uses the
gauge covariant vertex operators introduced in \cite{bsfc} and gives
manifestly gauge covariant equations.  The calculation is a little more
tedious than the first method, but it,too, gives the entire momentum
dependence.

            The gauge covariant method has been used to derive the
Yang-Mills equations\cite{bsfc}.  It is appropriate then to ask whether
it can be used
to calculate higher order corrections just as in the Abelian case.  As a
test of the method
we calculate higher derivative corrections to Yang Mills
equation and check that the result is covariant.

This paper is organized as follows:  In Section II we use use the gauge
fixed version of the proper time equation to calculate the leading
corrections to Maxwell's equation and explain the difference
between this calculation and a $\beta$-function calculation.
  In Sec III we do
this in the
gauge covariant method.  In Section IV we summarize the results of
a calculation of higher order corrections to the Yang-Mills equation.
We conclude in Section V with some comments.

\newpage
\section{Gauge Fixed Proper Time Equation}
\setcounter{equation}{0}

The proper time equation is
\be   \label{PT}
\frac{d}{dlnz} z^{2} <i \kim \p _{z} X^{\mu} \e i l_{1} ^{\nu} \p _{w}
X^{\nu} e^{i l_{0}. X(0)}>\, =\, 0
\ee
If we extract the coefficient of $\kim$ in the above equation we get
$\frac{\delta {\cal L}}{\delta A_{\mu}}$.
The expectation value uses the measure
\be
\int {\cal D} X exp \{ i \int _{M}
d^{2}z \p _{z} X^{\mu} \p _{\bar{z}} X^{\mu} +
\int _{\partial M}A^{\mu}(X) \p _{z} X^{\mu} dz  \}
\ee
Here $A_{\mu} (X) = \int d\ko A_{\mu} (k_{0}) \e $ is the background
Abelian gauge field.  We have used the notation $A^{\mu} (\ko )
\equiv \kim \,\, ;\, \, A^{\rho} (\qo ) \equiv \qir \, ...etc.$ here and
below.

The first non-zero correction to Maxwell's equation
is due to two insertions of the
$A_{\mu}$ vertex operator.
Thus we have to calculate:
\be    \label{KPQL}
\frac{d}{dlnz} z^{2} \int _{0}^{z} du \int_{0}^{u} dv
\ee
\[
<
ik_{1}^{\mu} \p _{z} X^{\mu} e^{ik_{0}.X(z)}
ip_{1}^{\rho} \p _{u} X^{\rho} e^{ip_{0}.X(u)}
iq_{1}^{\sigma} \p _{v} X^{\sigma} e^{iq_{0}.X(v)}
il_{1}^{\nu} \p _{w} X^{\nu} e^{il_{0}.X(0)} >
\]
\[
=\, \, 0
\]
We have chosen a particular ordering of momenta here.  In an
S-matrix calculation we would have to sum over all possible orderings,
but here, since the momentum is an integration variable,
we do not have to
worry about that.  The vertex operators are assumed to
satisfy the physical
state conditions, namely,
\be \label{PHS}
\ko ^{2} = \ko .\ki =0
\ee
There are various possible contractions that can be made in (\ref{KPQL}).
We will concentrate on those that give rise to terms of the form
$(\ki . p_{1} \qi . l_{1})$,  $ (\ki . \qi p_{1} . l_{1})$ or
$(\ki . l_{1} p_{1} . \qi) $.

Thus, consider the first one, viz., $\ki . p_{1} \qi . l_{1}$ :
\be     \label{IV}
\frac{d}{dlnz} z^{2} \int _{0}^{z} du \int_{0}^{u} dv
\ee
\[
\frac{1}{(z-u)^{2}}
\frac{1}{v^{2}}
(z-u)^{k_{0}.p_{0}}
(z-v)^{k_{0}.q_{0}}
(z)^{k_{0}.l_{0}}
(u-v)^{p_{0}.q_{0}}
(u)^{p_{0}.l_{0}}
(v)^{q_{0}.l_{0}}
\]
The integral becomes:
\be               \label{V}
(z)^{k_{0}.l_{0}} \int _{0}^{z} du
(z-u)^{k_{0}.p_{0}-2}
(u)^{p_{0}.l_{0}+q_{0}.l_{0}+p_{0}.q_{0}-1}
\ee
\[
\int _{0} ^{1} dv'
(z-v)^{k_{0}.q_{0}}(1-v')^{p_{0}.q_{0}}(v')^{q_{0}.l_{0}-2}
\]
Here $ v' \equiv v/u $.  We can expand $(z-v) ^{k_{0}.q_{0}}$ in a
powers of $\ko . \qo $ (if it is small).
The lowest term is just 1, and we get:
\be                     \label{VI}
(z)^{-2} \int _{0}^{1} du'
(1-u')^{k_{0}.p_{0}-2}
(u')^{-1+\delta}
\int _{0} ^{1} dv'
(1-v')^{p_{0}.q_{0}}(v')^{q_{0}.l_{0}-2}
\ee
We have used $p.l +q.l + p.q \equiv \delta $( $\approx \,  0$
 for on-shell photons) and $ u'\equiv \frac{u}{z}$. (We have
dropped the subscript `$0$' on the momentum variables in some
of the equations.)
 Naively,
acting on (\ref{VI}) with $\frac{d}{dln z} z^{2}$ would give
zero since there is no $z$-dependence in the integrals.
 However, as shown in \cite{bspt} one has to regulate the integrals
in (\ref{IV}) by modifying the limits to
$\int _{a} ^{z-a} \int _{a} ^{u-a}$ where $a$ is a short distance
cutoff on the world sheet.
Thus (\ref{VI}) becomes:
\be                     \label{VII}
\int _{a}^{1-\frac{a}{z}} du'
(1-u')^{k_{0}.p_{0}-2}
(u')^{-1+\delta}
\int _{\frac{a}{u}}^{1-\frac{a}{u}} dv'
(1-v')^{p_{0}.q_{0}}(v')^{q_{0}.l_{0}-2}
\ee
If we assume that $p.q \approx 0$, we need not regulate the upper
limit of the integral.  The second integral can be expanded in
a power series in $a/u$ to give:
\be \label{VIII}
B(-1+q.l,p.q+1) -
\frac{(a/u)^{q.l -1}}{q.l -1} -
\frac{(a/u)^{q.l }}{q.l }p.q +...
\ee
In (\ref{VIII}) we have kept only the terms that diverge as
$a \rightarrow 0$.
The linear divergence is due to tachyon exchange and we subtract it
 by
hand.
   If we insert (\ref{VIII})
    into (\ref{VII}) and do the $u'$-integral we get
for the $z$-dependent part
\be      \label{X}
\frac{(a/z)^{k.p-1}}{k.p-1} +
\frac{(a/z)^{k.p}}{k.p} +
\frac{(a/z)^{\delta}}{\delta}
\ee
and it multiplies
\be     \label{IX}
B(-1+q.l,p.q+1) -\frac{p.q}{q.l}
\ee
This is nothing but the $S$-matrix with its (massless) poles
subtracted.  Expanding(\ref{X}) in powers of $k.p \, \approx \, \delta
\, \approx \, 0$, we see that
the coefficient of $ln z$ is 2.
Thus, as argued in \cite{bspt}, the proper-time equation is
guaranteed to give the effective equations as determined from the
$S$-matrix.  Of course in the case of photon-photon scattering
there are no massless poles in any channel since there is no three-
photon vertex.  In fact, we have verified explicitly that
when contributions from all the terms
involving permutations of the momenta are added, the poles cancel.
In the case of Yang-Mills theory the poles will survive in the
$S$-matrix, but will still be subtracted out in the proper-time
equation, exactly as in the present case,  (\ref{IX}).  If we
expand (\ref{IX}) we get
\be     \label{XI}
\frac{1}{q.l-1}[\frac{p.q}{q.l} +1][1-(q.l)(p.q)\zeta (2)] -
\frac{p.q}{q.l} + higher \, order...
\ee
($\zeta (2) = \frac{\pi ^{2}}{6}$)
For $q.l \approx p.q \approx 0$ we thus get
\be    \label{XII}
-1 -2(k.q)(p.q)\zeta (2)
\ee
Thus, finally,
the contribution to the equation is a term of the form:
\be         \label{XIII}
\ki . p_{1} \qi . l_{1} (k.q)(p.q) = \frac{\ki . p_{1} \qi . l_{1}}
{2} [(l.q)^{2} - (p.q)^{2} - (k.q)^{2} ]
\ee
which comes from a term
\be        \label{XIV}
[F^{4} - 1/4 (F^{2})^{2}]
\ee
which is precisely the quartic piece of the Born-Infeld action.
Note that the $-1$ in  (\ref{XII})  did not get subtracted
because it did not come from a logarithmic divergence.  Nevertheless
when all the permutations are added it drops out, just as in the
$S$-matrix calculation. (Of course, we must remember that the
proper-time equation is an equation of motion.  Thus the
contribution to the equation of motion that we have obtained
is really the part that multiplies $ \kim $ in (\ref{XIII}).  This
obviously corresponds,
upto an overall combinatoric factor,
to varying w.r.t $A_{\mu}$ in (\ref{XIV}).)

Let us now turn to the $\beta $-function calculation of \cite{ac}.
The cubic term in the
$\beta$-function is
\be \label{XV}
(F^{2})_{\lambda \nu} \p ^{\nu}F^{\lambda}_{\mu}
\ee
It can easily be checked that it differs from the full equation
(at this order) by terms proportional to $\mup F_{\mu \nu}$,
i.e. Maxwell's equation.
Thus, in particular, the $(F^{2})^{2}$ term (more precisely, its
variation) is not contained in the $\beta$-function.  It is easy to
see why.
In a $\beta$-function calculation,
 Maxwell's equation is the coefficient of the
logarithmic divergence in the lowest order graph.
In a higher order graph, a lower order
divergence is necessarily cancelled by a counter-term
in any renormalization scheme.  Thus at the cubic order the
$\beta $-function will not have anything proportional to Maxwell's
equation.  The proper-time equation, on the other hand, picks out
a logarithmic divergence.\footnote{It actually looks for $ln z$
pieces.  But $ln z$ always occurs as $ln (z/a)$, where $a$ is the
short distance cutoff,
for dimensional
reasons, so the two procedures are
equivalent.} - it does not matter whether it is
from a divergent subgraph or an overall divergence.  Thus
terms proportional to $\mup F^{\mu \nu}$ would show up in a
higher order calculation.
The only
subtractions are $(log) ^{2}$ (or higher)
divergences - which corresponds to
subtracting pole terms multiplying the $ln z$ piece.  This is
exactly the procedure for determining the effective action from
an $S$-matrix.  The above arguments thus show why
the $\beta$ function cannot be equal to the equation of motion,
and also  why it is plausible that the proper-time
equation is the same as the equation of motion.  In the case
of the tachyon it was shown in \cite{bspt} that the two are indeed
equivalent to all orders.  In the present case we have verified it
only to cubic order (in the field strength, but to all orders
in derivatives).  However, the nature of the argument in \cite{bspt}
did not depend in
any specific way on the properties of the tachyon and therefore
we expect
it to go through for all the modes.

Finally, it is important to note that (\ref{IX}) has the exact
momentum dependence (except for the restriction $k.q \approx \, 0 \,$).
Thus, this method gives {\em all the higher derivative corrections}
to the low energy equations of motion.  This is very difficult
to do in a $\beta $-function calculation.

\newpage
\section{Gauge Covariant Proper Time Equation}
\setcounter{equation}{0}

We now repeat this calculation in a manifestly covariant way
using the techniques of \cite{bsfc}.  We use the following
identity derived in \cite{bsfc}:
\be     \label{ID}
\e i \ki \p X = \int _{0}^{1} d \al \p (e^{i\alpha \ko X} i \ki
X) + \int _{0}^{1} d \al [ e^{i\alpha \ko  X} X^{\nu} \p X^{\mu}
(i)^{2} \al \ko ^{[\nu} \ki ^{\mu ]} ]
\ee
and insert in the proper time equation used there:
\be       \label{3.2}
\frac{d}{d \Sigma} \int d z_{1} \int d z_{2}
<  i\kim \p _{z_{1}} X ^{\mu} e^{ik_{0}X}
  i l_{1}^{\nu} \p _{z_{2}} X ^{\nu} e^{il_{0}X} >
\ee
As explained in \cite{bsfc,bsg}, the integrals over $ z_{1}, z_{2}$
allow us to throw away total divergences and thereby makes the result
gauge invariant.  This calculation is similar
to that in \cite{bsfc} where we calculated the cubic and quartic terms
in the Yang-Mills coupling.  Comparing with that calculation, it is easy
to see that only the second term on the RHS of (\ref{ID})
contributes in the Abelian case.\footnote{The other terms
contribute to the cubic and quartic Yang-Mills coupling.}
  Thus to lowest order we just have
Maxwell's equation:
\be    \label{3.3}
\int dz_{1} \int dz_{2} \int d \alpha \alpha \int d \beta \beta
k_{0}^{[\mu}k_{1}^{\nu ]} l_{0}^{[\rho }l_{1}^{\sigma ]}
<  X^{\mu} \p _{z_{1}} X ^{\nu} e^{i\alpha k_{0}X}
  X^{\rho} \p _{z_{2}} X ^{\sigma} e^{i\beta l_{0}X} >
\ee
\be      \label{3.4}
= \, 1/4\int dz_{1} \int dz_{2}
k_{0}^{[\mu}k_{1}^{\nu ]} l_{0}^{[\rho }l_{1}^{\sigma ]}
(\delta ^{\mu \rho} \delta ^{\nu \sigma} \Sigma \p _{z_{1}} \p_{z_{2}}
\Sigma                                                      +
\delta ^{\mu \sigma} \delta ^{\nu \rho}  \p _{z_{1}} \Sigma \p_{z_{2}}
\Sigma )
\ee
where  $\Sigma = ln(z_{1}-z_{2})$.
If we now follow the procedure of \cite{bsfc} and vary w.r.t.
$\Sigma$, we get
\be    \label{3.5}
1/2\int dz_{1} \int dz_{2}
k_{0}^{[\mu}k_{1}^{\nu ]} l_{0}^{[\rho }l_{1}^{\sigma ]}
(\delta ^{\mu \rho} \delta ^{\nu \sigma} -
\delta ^{\mu \sigma} \delta ^{\nu \rho}) \p _{z_{1}}  \p_{z_{2}}
\Sigma   =0
\ee
  The coefficient of $\kim$ is Maxwell's equation: $\mup F^{\mu \nu}
=0$.

One can also follow the technique \cite{bspt} of looking at the
logarithmic deviation from $\frac{1}{(z_{1}-z_{2})^{2}}$.  Thus we
operate on (\ref{3.5}) with $\frac{d}{dln(z_{1}-z_{2})} (z_{1}-z_{2})
^{2}$.  In this
case we do not integrate over $z_{1},z_{2}$.  We then get the same
result:
\be      \label{3.6}
1/4 k_{0}^{[\mu}k_{1}^{\nu ]} l_{0}^{[\mu }l_{1}^{\nu ]}
\frac{d}{d ln(z_{1}-z_{2})}(ln(z_{1}-z_{2}) + 1)
\ee
\be      \label{3.7}
= \, 1/4 k_{0}^{[\mu}k_{1}^{\nu ]} l_{0}^{[\mu }l_{1}^{\nu ]}
\ee
The coefficient of $\kim$ gives Maxwell's equation as before, except
that the coefficient is 1/4 rather than 1.\footnote{This discrepancy
is not important for the purposes of this paper.  However it is an
issue that needs to be resolved.}

We now look at the next
non-trivial order - namely quartic, since the cubic term is easily
shown to vanish.
  Our aim is to show explicitly
that the proper-time equation produces the contribution of the
$(F^{2})^{2}$ term in the Born-Infeld action.  To this end we consider
the following:
\[
\int dz_{2} \int dz_{3} \int d \alpha \alpha  d \beta \beta
d\gamma \gamma d\delta \delta
\]
\[
<  X^{\mu _{1}} \p _{z_{1}} X ^{\nu _{1}} e^{i\alpha k_{0}X}
  X^{\mu _{2}} \p _{z_{2}} X ^{\nu _{2}} e^{i\beta p_{0}X}
  X^{\mu _{3}} \p _{z_{3}} X ^{\nu _{3}} e^{i\gamma k_{0}X}
  X^{\mu _{4}} \p _{z_{4}} X ^{\nu _{4}} e^{i\delta p_{0}X} >
\]
\be   \label{3.8}
 k_{0}^{[\mu _{1}}k_{1}^{\nu _{1}]} p_{0}^{[\mu _{2}}p_{1}^{\nu _{2} ]}
\ee
We consider a specific contraction of the type
$
\delta ^{\mu _{1} \mu _{2}}
\delta ^{\nu _{1} \nu _{2}}
\delta ^{\mu _{3} \mu _{4}}
\delta ^{\nu _{3} \nu _{4}}
 $.  We get the proper time equation:
\[
\frac{d}{d ln(z_{1})}  (z_{1})^{2}
F^{2}F^{2} \int _{0}^{z_{1}} dz_{2} \int _{0}^{z_{2}}dz_{3}
\]
\[
[\frac{ln(z_{1}-z_{2})+1}{(z_{1}-z_{2})^{2}}]
[\frac{lnz_{3}+1}{z_{3}^{2}}]
\]
\be   \label{3.9}
(z_{1}-z_{2})^{\alpha \beta k_{0} p_{0}}
(z_{1}-z_{3})^{\alpha \gamma k_{0} q_{0}}
(z_{2}-z_{3})^{\beta  \gamma p_{0} p_{0}}
(z_{2}-z_{4})^{\beta \delta p_{0} l_{0}}
(z_{3}-z_{4})^{\gamma \delta q_{0} l_{0}}
(z_{1})^{\alpha \delta k_{0}l_{0}}
\ee
\[
= \, \, 0
\]
There is a new momentum conservation equation:
\be   \label{3.10}
\alpha \ko + \beta p_{0}+ \gamma \qo +\delta l_{0} =0
\ee
that constrains $\alpha , \beta , \gamma , \delta $.  For simplicity
we will refer to $\alpha \ko $ as $k$  etc. and restore these factors at
the end.
We rewrite (\ref{3.9}) as:
\be      \label{3.11}
\frac{d}{dln z_{1}}z_{1}^{2} z_{1}^{k.l} \int dz_{2}
(z_{1}-z_{2})^{-2+k.p + \nu} z_{2}^{p.l} \int d z_{3}
 z_{3}^{-2+q.l + \mu} (z_{2}-z_{3})^{p.q}(z_{2}-z_{3})^{k.q} \,
= \, 0
\ee
If we expand (\ref{3.11}) in powers of $\nu , \mu $, they multiply
powers of $ln (z_{1} - z_{2})$ and $ln z_{3}$ respectively.  Thus
we evaluate (\ref{3.11}) for general $\mu , \nu $ , expand the
result in powers of $\mu , \nu$, and look at the $\mu ,\nu $-
independent terms, those
linear in $\mu , \nu $ and those bilinear in $\mu , \nu $.
The sum of these terms (with $\mu , \nu $ set to 1)
will give us the result of the integral (\ref{3.9}).  To simplify
the integral we set $k.q =0$.  The integral gives:
\be      \label{3.12}
z_{1}^{\mu +\nu}B(-1+k.p + \nu ,\mu ) B(-1+q.l+\mu,1+p.q)
\ee
We should actually regularize the integral, but since it just has the
effect of subtracting the poles, we will do it by hand at the end.
Using the expansion
\[
\frac{\Gamma (1+x) \Gamma (1+y)}{\Gamma (1+x+y)} = 1 - xy \zeta (2)
+ .....
\]
we get for the coefficient of $ln z_{1}$:
\[
(\mu + \nu ) \{ (1+ \frac{\mu}{k.p + \nu -1})(1+ \frac{\mu}{k.p + \nu})
\frac{1}{\mu}[1- \mu (k.p + \nu ) \zeta (2)]\}
\]
\be      \label{3.13}
\{\frac{1}{\mu +q.l -1}[1+\frac{p.q}{q.l+\mu}]
[1-(q.l+\mu )p.q\zeta (2)]\}
\ee
The pole terms are to be subtracted as usual.  If we look at the term
multiplying $\zeta (2)$ we find:
\be     \label{3.14}
\int d \alpha \alpha
\int d \beta \beta \int d \gamma \gamma
\int d \delta \delta [-\frac{\mu \nu}
{q.l-1} \zeta (2) ] = \frac{1}{16} \mu\nu \zeta (2) +
\,\, higher \,\, order \,\, in \,\, q.l
\ee
Thus (\ref{3.9}) becomes $\frac{1}{16}\mu \nu \zeta (2) F^{2}F^{2}$.
Thus we see that there is an $F^{2}F^{2}$ with some non-zero
coefficient , which is what we wanted to show.

This calculation is a little more tedious than the $\beta$-function
calculation.  But as mentioned in the previous section,
we should realize that we have a result that is
{\em valid to all orders in momenta}.
All we have to do is to expand the
Beta function in (\ref{3.12}) in powers of momenta and do the
$\alpha , \beta , \gamma , \delta $ integrals (note that
$k$ in (\ref{3.12}) stands for $\alpha \ko $ and so on). The only
approximation that has been made is that we have set $k.q =0$, thus
our result is correct to lowest order in $k.q$.

\newpage
\section{Higher Order Corrections in Yang-Mills Theories}
\setcounter{equation}{0}

As another application of the gauge covariant proper-time
formalism of \cite{bspt,bsg,bsfc} we have calculated higher
derivative corrections to the Yang-Mills equation - corrections
of the form $Tr F^{3}$ or $DFDF$\cite{do}.
More precisely, we have calculated the
 terms involving three $A$ fields and three derivatives (The term
involving four derivatives and two A fields is trivial to calculate),
and verified explicitly that the result is covariant.  The calculation
is straightforward, but a little tedious, and we will only give an
outline and some intermediate results.  Our main purpose is to
test the method.

We start with the following:
\be   \label{4.1}
Tr[T^{a}T^{b}T^{c}]\int dz \int _{w}^{z} dz_{1}\int d w
<i\ki ^{a} \p_{z} X \e i\qi ^{b} \p _{z_{1}}X \qe i p_{1}^{c}
\p_{w} X e^{i p_{0}X(w)} >
\ee
 We replace each vertex operator with the expression on the RHS
of (\ref{ID}), reproduced here for convenience:
\be       \label{4.2}
\e i \ki \p X = \underbrace{
\int _{0}^{1} d \al \p (e^{i\alpha \ko X} i \ki
X)}_{a}
+ \underbrace{
\int _{0}^{1} d \al [ e^{i\alpha \ko  X} X^{\nu} \p X^{\mu}
(i)^{2} \al \ko ^{[\nu} \ki ^{\mu ]}]}_{b}
\ee
We will refer to the total derivative piece as `a' and the ``gauge
invariant '' piece as `b'.  Substituting (4.2) into (4.1) gives
eight terms that we can conveniently label as : `aaa', `aba',`aab' ...
etc.  The antisymmetry of the terms will ensure that $Tr[T^{a}T^{b}
T^{c}]$ is multiplied by an antisymmetric (in$ a,b,c$) term, thus
converting it to $f^{abc}$.  The `aaa' term is easily seen to be zero.
The `aab' term gives (suppressing group theory indices)
\be  \label{4.3}
-<e^{i\alpha (k_{0} + k_{1}) X(z)}\p_{z} e^{i \beta (q_{0} +q_{1})
X(z)}i^{2} \gamma ^{2} X^{\sigma }\p_{w}X^{\theta} p_{0}^{[\sigma}
p_{1}^{\theta ]}e^{i\gamma p_{0} X(w) }>
\ee
We have written $\ki$ and $\qi$ in the exponent with the understanding
that
we are to keep only the piece linear in $\ki \qi $.  This gives, for
the piece cubic in momentum,
\be        \label{4.4}
\int dw \int dz \gamma ^{2} \alpha \beta p_{0}^{[\sigma}
p_{1}^{\theta ]} \{
 [(\ko + \ki ) ^{\sigma} ( \qo + \qi )^{\theta }
\Sigma \p _{z} \p_{w} \Sigma \Sigma (-1)
[\alpha (\ko + \ki ) + \beta (\qo + \qi )].\gamma p_{0}]
\ee
\[
+[(\ko + \ki ) ^{\theta} ( \qo + \qi )^{\sigma }
\p _{z} \Sigma \p_{w} \Sigma \Sigma (-1)
[\alpha (\ko + \ki ) + \beta (\qo + \qi )].\gamma p_{0}]  \}
\]
Here, as always $\Sigma = ln (z-w)$.  If we multiply by
$(z-w)^{2}$ and pick the coefficient of $ln (z-w)$  we are
left with the second term ($\p _{w} \Sigma \p _{z} \Sigma \Sigma $).
On the other hand, treating $\Sigma $ as a field \cite{bsg,bsfc},
 if we allow integration by parts, we get contribution
from both terms.  Thus depending on which procedure we use, we get
different numerical coefficients.  This is the same ambiguity
encountered in Sec III.
Fortunately, all terms at this order
have the same form, so we can consistently pick one procedure.
Presumably, comparison with higher and lower order terms will allow
us to decide which is the right procedure.  For this calculation
we allow ourselves to integrate by parts to bring everything into
the form $ \Sigma \p _{z} \p _{w} \Sigma $.  This gives, after doing
the $\alpha , \beta , \gamma $ integrals: (for`aab')
\be     \label{4.5}
1/4p_{0} ^{[\sigma} p_{1}^{\theta ]} \{ \ki^{[\sigma}\qi ^{\theta ]}
(\ko + \qo ).p_{0} + \ko ^{[\sigma }\qi ^{\theta ]}\ki . p_{0}
+\ki ^{[\sigma }\qo ^{\theta ]} \qi p_{0} \}
\ee
The other terms `aba' and `baa' are obtained by cyclic permutations.
A similar calculation gives for the `abb' term
\be  \label{4.6}
-3/4\ki .(\qo - p_{0} ) \qo ^{[ \mu} \qi ^{\nu ]} p_{0}^{[ \mu  }
p_{1}^{\nu ]}
\ee
Adding (\ref{4.5}) and (\ref{4.6}) gives (we have suppressed
the group indices of $k,q,p$ above)
\be    \label{4.7}
\{ \p _{\sigma } ( \mup A_{\nu} - \nup A_{\mu} ) \p ^{\sigma }
[A_{\mu} , A_{\nu}] + \p_{\sigma } (\mup A_{\nu} - \nup A_{\mu})
[A^{\sigma}, (\mup A_{\nu} - \nup A_{\mu}]\}
\ee
with $A_{\mu} = A_{\mu} ^{a} T^{a} $ and $tr\{ T^{a},[T^{a},T^{c}] \}
= f^{abc}$.(\ref{4.7}) is obviously the cubic piece in
$Tr \{ D_{\sigma } F^{\mu \nu}
D^{\sigma } F_{\mu \nu}\}$.  Finally the `bbb' term gives, upto an
overall constant, $Tr[F^{3}]$.
This verifies the gauge covariance of the corrections to Yang-Mills'
equations.

It would be interesting to see whether the concept of ``covariant
derivative'' can be fruitfully introduced in such calculations to
simplify the algebra.

\newpage
\section{Conclusions}
We have compared the proper-time formalism with the $\beta $ -function
method.  We have seen that the proper-time formalism gives the full
equation of motion, including the prefactor, known usually as the
Zamolodchikov metric.  In the case of Abelian gauge fields we showed
explicitly that the $(F^{2})^{2}$ term is obtained.  We did this in two
different ways - one being gauge fixed and very similar to the
$S$-matrix calculation.  The other method is gauge covariant.
It is important to point out that in both cases one can get results
to arbitrary high order in momenta with very little work.  The
reason it is easier than calculating the $\beta$-function
is that one does not have to worry about subtracting lower
order divergences.  It may in fact turn out that an easier way to
calculate the $\beta $- function is to calculate the proper time equation
and divide by the Zamolodchikov metric.

In Section IV we also presented results regarding higher order
(derivative) corrections in Yang-Mills theory using the covariant
proper-time method.  We thus verified that the method is consistent
and gives gauge covariant results.

In conclusion, the proper time method seems promising as a means of
calculating low energy equations of motion in a gauge covariant way.
It can also be extended off-shell as in \cite{bsfc} by introducing
a finite world sheet cutoff.  The massive modes also appear in a natural
way there.  We hope that by considering both the low energy
and high energy sytems by means of the same equation one can interpolate
smoothly between them.  This should provide some insight into the
symmetries of strings and the role of the massive modes.

\end{document}